\documentclass[journal=jacsat,manuscript=article]{achemso}
\SectionsOn
\SectionNumbersOn
\usepackage[version=3]{mhchem} 
\usepackage{xspace}
\usepackage{soul}
\usepackage[normalem]{ulem}
\usepackage{xcolor}
\usepackage{hyperref}
\usepackage{amsfonts}
\usepackage{amsmath}
\DeclareMathOperator*{\argmax}{arg\,max}

\DeclareCaptionLabelFormat{myformat}{#1~S#2}
\author{Hanfeng Zhai}
\affiliation[Cornell University]{Sibley School of Mechanical and Aerospace Engineering, \\ Cornell University, Ithaca, NY 14850, USA}
\author{Jingjie Yeo}
\email{jingjieyeo@cornell.edu}
\affiliation[Cornell University]{Sibley School of Mechanical and Aerospace Engineering, \\ Cornell University, Ithaca, NY 14850, USA}

\title[Bayesian optimization for antibiotic surfaces]
  {Computational design of antimicrobial active surfaces via automated Bayesian optimization}%

\abbreviations{IR,NMR,UV}
\keywords{American Chemical Society, \LaTeX}

\begin{document}







\begin{abstract}
Biofilms pose significant problems for engineers in diverse fields, such as marine science, bioenergy, and biomedicine, where effective biofilm control is a long-term goal. The adhesion and surface mechanics of biofilms play crucial roles in generating and removing biofilm. Designing customized nano-surfaces with different surface topologies can alter the adhesive properties to remove biofilms more easily and greatly improve long-term biofilm control. To rapidly design such topologies, we employ individual-based modeling and Bayesian optimization to automate the design process and generate different active surfaces for effective biofilm removal. Our framework successfully generated ideal nano-surfaces for biofilm removal through applied shear and vibration. Densely distributed short pillar topography is the optimal geometry to prevent biofilm formation. Under fluidic shearing, the optimal topography is to sparsely distribute tall, slim, pillar-like structures. When subjected to either vertical or lateral vibrations, thick trapezoidal cones are found to be optimal. Optimizing the vibrational loading indicates a small vibration magnitude with relatively low frequencies is more efficient in removing biofilm. Our results provide insights into various engineering fields that require surface-mediated biofilm control. Our framework can also be applied to more general materials design and optimization.
\end{abstract}

{\textbf{\em Keywords:} Biomaterials; Bayesian optimization; machine learning; biofilms; microstructure; individual-based modeling}
\section{Introduction}

Biofilms and biofouling are significant threats to food and health systems as reported by the U.S. Environmental Protection Agency\cite{eps_biofilm_threat}. Moreover, the formation and attachment of biofilms pose serious problems for marine engineering\cite{marine, marine2} and biomedical treatments\cite{biomed_1, biomed_2}, where long-term biofilm control is desired. For example, biofilms adhering to medical implant devices lead to infections \cite{biofilm_book_intro}. Biofilms also potentially lead to medical treatment failures like ventilator-associated pneumonia, eye infection, and urinary tract infections \cite{biofilm_threat_review}. 
Furthermore, the World Health Organization recently reported that antimicrobial resistance is becoming a grave issue that requires immediate action\cite{who}, indicating that the overuse of chemical treatments may not be an ideal roadmap for long-term biofilm control. Hence, environmentally benign and sustainable biofilm control strategies are urgently needed to prevent treatment failure caused by biofilm resistance. From the perspective of biomechanics, the adhesion between bacteria cells and attached surfaces plays a critical role in the formation and maturation of the biofilms\cite{adhesion1, adhesion2}. Therefore, a promising method to attenuate adhesion is to tune the surface properties and engineer antifouling materials that resist bacterial colonization\cite{topo1, topo2}.

Engineering surface properties for biofilm control has been of interest for decades. However, successful methods like tar paints and copper panel sidings tend to leach biocides \cite{fouling_history}, which leads us to the question: are there environmentally friendly approaches to tackle such biofilm issues? Two approaches were proposed to diminish biofilm's adhesive properties: (1) tailoring chemical properties at the molecular level, with a specific focus on polymeric design\cite{ober_review}, and (2) altering the topographies of the active surfaces to tune the nano- and micro-mechanical properties \cite{topo1}. From the chemical perspective, Zhang et al.\cite{chem_biofilm} tailored the polar functionalities of PDMS to design polymeric antifouling surfaces. Xu et al.\cite{chem2_biofilm} used highly hydrophilic sulfoxide polymers to produce antifouling polymer brushes. Besides elastomers and polymer brushes, block copolymers, hydrogels, and other materials can also be utilized for antibiofilm designs\cite{ober_review}. Nonetheless, tailoring polymeric properties requires precise chemical operations at the molecular level which are expensive and not scalable for bulk manufacturing at the current stage. In contrast, due to advances in additive manufacturing, altering surface nanotopographies are much cheaper, faster, and industrially scalable. Friedlander et al.\cite{pnas_activesurface} optimized the nanotopography by introducing submicrometer crevices. Hizal et al.\cite{acs_german_nanoengineer} showed that active surfaces reduce the adhesion of biofilms on such surfaces for ease of removal. The reduced contact area with the active surface topology with the biofilms leads to such reduced adhesion.
Both Bhattacharjee et al.\cite{change_surface_change_biofilm_growth} and Lohmann et al.\cite{activesurface} showed that by altering the topological parameters such as radii, height, and distances between the cones, the active surfaces exhibit different effects on biofilm growth. For instance, Bhattacharjee et al.'s work indicates that more densely distributed pillars with smaller radii can kill bacteria more efficiently. These studies consequently pose an important question: can these active surfaces be designed systematically to resist or promote biofilm formation under different physical environments? 


Designing active surfaces using many cones of tunable radii and heights as the methods proposed by Bhattacharjee et al. \cite{change_surface_change_biofilm_growth} and others\cite{pnas_activesurface, activesurface}, involves an infinite design space that is extremely challenging to explore comprehensively using either computational simulations or physical experiments. Due to advances in machine learning algorithms, heuristic optimization-based materials design methods may be a potential solution to tackle this tough question by exploiting sparse data points\cite{heuristic1, heuristic2}. Specifically, due to the heuristic characteristics and capacity for handling black-box functions, Bayesian optimization (BO) has been widely used in materials design\cite{Frazier_bayesian-optimization_materials-design}. Here, we employ BO as a toolkit to sample the large design space for optimizing the nanotopography.

Another common barrier encountered in studies of biofilms is lengthy experimental procedures that can take weeks to culture mature biofilm \cite{biofilm_time}. To bypass this barrier, digital twins, specifically multiscale computational modeling, may help to speed up design optimizations. Various methods have been proposed for modeling biofilm, spanning the molecular to the continuum scale. In particular, individual-based (a.k.a agent-based) modeling (IBM) is a rapidly maturing technique for simulating biofilm's multiscale and multiphysics characteristics\cite{ dem_nature_biofim_modeling,biofilm_ibm}. 

In this study, we aim to provide a fully digital, automated machine learning workflow for designing antimicrobial active surfaces for biofilm control. The workflow is based on coupled BO and the IBM platform from Newcastle University Frontier in Engineering Biology (NUFEB)\cite{Li2019} implemented in \texttt{LAMMPS}\cite{lammps}. We explore two typical scenarios of biofilm growth for nanotopology optimization: (1) biofilm growth under static conditions for designing active surfaces that resist biofilm growth, (2) biofilm subjected to constant shear flow for designing active surfaces under fluidic flow, such as the environments typically encountered in marine applications\cite{ober_review} and wastewater treatments\cite{bioenergy}. Inspired by recent work demonstrating that vibration may be a viable approach for removing biofilm\cite{vibration_teng_syracuse}, we propose another two scenarios of (3) vertical vibration and (4) lateral vibration of the active surface for biofilm detachment.

\begin{figure}[htpb]
    \centering
    \includegraphics[scale=0.55]{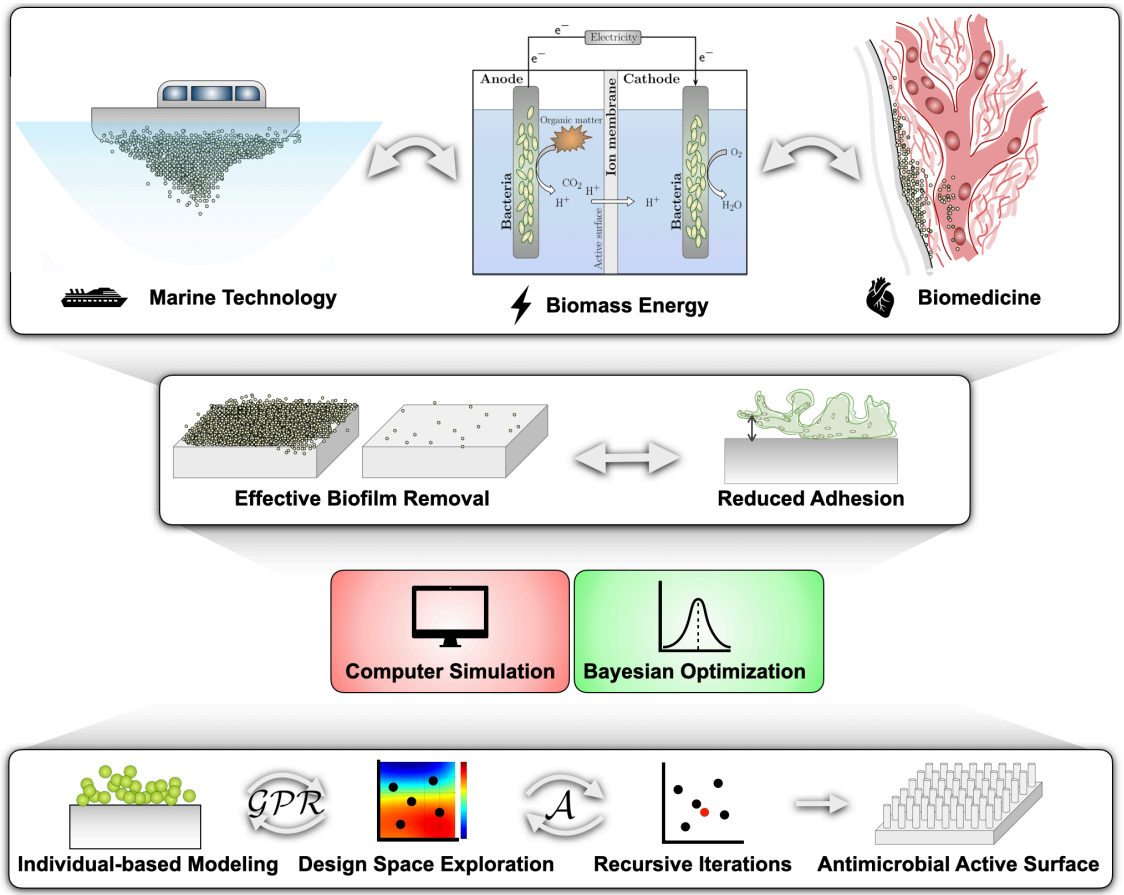}
    \caption{A representative schematic of this study. Biofilm and related issues plague many aspects of engineering but tuning the adhesive properties of surfaces by changing the surface topology can be a potential solution. To optimize these topologies, we use a machine-learned materials design workflow driven by individual-based simulations and BO. We construct a surrogate model via Gaussian process regression ($\mathcal{GPR}$) and iterative data search via the acquisition function ($\mathcal{A}$) to eventually propose the optimized active surfaces.}
    \label{general_schematic}
\end{figure}

 The paper is arranged as follows (Figure \ref{general_schematic}): In Section \ref{method} we introduce the computational methods of individual-based models and simulations (Sec. \ref{dem}) and the method of BO (Sec. \ref{bayesian_optimization}) using Gaussian process regression and an acquisition function in Sections \ref{gpr} and \ref{acquisition}, respectively. We then propose the basic simulation setup implemented in the Large-scale Atomic/Molecular Massively Parallel Simulator (\texttt{LAMMPS}) in Section \ref{simulation}. The general optimization workflow is explained in Section \ref{workflow}. In Section \ref{result} we detail how we construct meta-models from the optimization (Sec. \ref{metamodel}), extract the optimized geometries (Sec. \ref{geometry}), and subsequent biomechanical analyses (Sec. \ref{biomechanics}). Eventually, some interesting conclusions are drawn and potential research directions are pointed in Section \ref{conclusion}, including high thin-pillar shaped nanosurfaces are more efficient in removing biofilms under shear flow, yet short thick cones are found to be more efficient for vibrational biofilm removal.

\section{Methodology\label{method}}
\subsection{Individual-based Simulations\label{dem}}
 In individual-based models based on the Newcastle University Frontiers in Engineering Biology (NUFEB) framework \cite{Li2019}, each bacteria cell is modeled as a spherical particle. Biofilms are formed by cell division and extrusion of extracellular polymeric substances (EPS).
The microbe growth and decay are described by the differential equation:\begin{equation}
    \frac{d m_i}{dt} = \xi_i m_i \label{equation_growth}
\end{equation}where $m_i$ is the biomass of the $i^{th}$ bacteria cells and $\xi_i$ is the growth rate. Here, we employ the Monod-based method\cite{Monod1949} to model microbial growth, in which the growth rate is determined by the Monod kinetic equation driven by the local concentration of nutrients. The substrate is modelled as fully rigid particles.

Since we are essentially interested in the adhesive property, we must model the mechanical interactions of the particles. The particles are mechanically relaxed using the individual-based approach, solved via Newton's equation \begin{equation}
    m_i \frac{d{\bf v}_i}{dt} = \mathbf{F}_{c,i} + \mathbf{F}_{a,i} + \mathbf{F}_{d,i} + ... ,\label{newton_grow}
\end{equation}where $m_i$ is the mass of a particle, and $\mathbf{v}_i$ is the velocity. The contact force $\mathbf{F}_{c,i}$ is a pair-wise force between particles to prevent overlapping based on Hooke's law \begin{equation}
    \mathbf{F}_{c,i} = \sum_{j=1}^{N_i} \left( K_\mathbb{N} \delta \mathbf{n}_{i,j} - m_{i,j} \gamma_\mathbb{N}  \mathbf{v}_{i,j} \right)\label{f1}\end{equation}where $N_i$ is the total number of neighboring particles of $i$, $K_\mathbb{N}$ is the elastic constant for normal contact, $\delta \mathbf{n}_{ij}$ is the overlap distance between the center of particle $i$ and its neighbour particle
$j$. $\gamma_\mathbb{N} $ is the viscoelastic damping constant for normal contact, and $v_{i,j}$ is the relative velocity of the two particles. The EPS adhesive force ${\bf F}_{a,i}$ is a pair-wise interaction modelled as a van der Waals force\begin{equation}
    \mathbf{F}_{a,i} = \sum_{j=1}^{N_i} \frac{H_a r_{i,j}}{12 h_{min,i,j}^2 } \mathbf{n}_{i,j}\label{f2}
\end{equation}where $H_a$ is the Hamaker coefficient, $r_{i,j}$ is the effective outer-radius of the $i^{th}$ and $j^{th}$ particles. $h_{min,i,j}$ is the minimum separation distance of the two particles, and $\mathbf{n}_{i,j}$ is the unit vector from particle $i$ to $j$. The drag force $\mathbf{F}_{d,i}$ due to fluid-particle interactions in fluid flow is determined from\begin{equation}
    \mathbf{F}_{d,i} = \frac{V_{p,i}}{\epsilon_{f,i}\epsilon_{s,i}} \beta_i (\mathbf{u}_{p,i} - \mathbf{U}_{f,i})\label{f3}
\end{equation}where $\epsilon_{s,i}$ is the particle volume fraction, $\epsilon_{f,i} = 1 - \epsilon_{s,i}$ is the fluid volume fraction, $V_{p,i}$ and $\mathbf{u}_{p,i}$ are volume and velocity of the $i^{th}$ particle, respectively. $\mathbf{U}_{f, i}$ is the fluid velocity imposed on particle $i$ and $\beta_i$ is the drag correction coefficient.

Mechanical equilibrium is achieved when the average pressure of the microbial community reaches a plateau. The average pressure of the system is\begin{equation}
    P = \frac{1}{3V} \left( \sum_{i=1}^N m_i \mathbf{v}_i \cdot \mathbf{v}_i + \sum_{i=1}^N\sum_{j>i}^N \mathbf{r}_{i,j} \cdot\mathbf{F}_{i,j} \right)\label{relax_equilibrium_mechanical}
\end{equation}where $V$ is the sum of the volumes of particles. The first term in the bracket is the contribution from the kinetic energy of each particle. The second term is the interaction energy, where $\mathbf{r}_{i,j}$ and $\mathbf{F}_{i,j}$ are the distance and force between two interacting particles $i$ and $j$, respectively.

In the simulations, shear flow is applied for biofilm removal, where fluid dynamics is coupled to the individual-based method. The hydrodynamics is incorporated in NUFEB via the two-way coupled CFD-DEM approach \cite{dem_cfd, Li2019}. The governing equations for the fluid phase are\cite{old_fluid_equation}
\begin{equation}
    \nabla \cdot \left(\epsilon_s \mathbf{U}_s + \epsilon_f \mathbf{U}_f \right) = 0\label{cfd1}
\end{equation}and \begin{equation}
    \frac{\partial (\epsilon_f \mathbf{U}_f)}{\partial t} + \nabla\cdot\left( \epsilon_f \mathbf{U}_f \mathbf{U}_f\right) = \frac{1}{\rho_f} \left(- \nabla P + \epsilon_f \nabla\cdot\mathcal{R} + \epsilon_f \rho_f \mathbf{g} + \mathbf{F}_f\right)\label{cfd2}
\end{equation}
where $\epsilon_s$, $\mathbf{U}_s$ and $\mathbf{F}_f$ are the solid volume fraction, velocity, and fluid-particle interaction forces of the bacteria, respectively. 

By coupling the bacteria growth dynamics with contact, adhesive, and drag forces, mechanical relaxation to equilibrium, and fluid dynamics, we are able to simulate the biofilm behavior on different surface topologies. Based on this IBM, the simulation details are further explained in Section \ref{simulation}.

\subsection{Bayesian Optimization\label{bayesian_optimization}}

The overall goal of the optimization process is to minimize or maximize an objective function, which in our case is the total bacteria cells after fluidic shear is applied to remove the biofilm. Using $y = f({\bf x, p})$ to denote a multivariate function relation, variables $\bf x$ and parameters $\bf p$ relates to the output $y$ through the function(al) form of $f$, where ${\bf x} = [R_x, R_y, ...]$ are the overall design parameters in the numerical simulation. The optimization process can be simplified as \begin{equation}
    \begin{aligned}
    \min y = \mathcal{N}_{BC} = f({\bf x, p}),\\
    {\rm subject\ to}\quad {\bf x}_{\rm LB} \leq \mathbf{x} \leq \mathbf{x}_{\rm UB},\ 0 \leq R_y \leq R_x \leq \frac{L_x}{n}\\
    \mathbf{x} = \left[R_x, R_y, h, n, (M, T)\right],\ \mathbf{p} = \left[\alpha, \xi, \mathbb{L}, \mathcal{T}, \mathcal{B}\right]
    \end{aligned}\label{opt_problem_equation}
\end{equation}

Here, $y = \mathcal{N}_{BC}$ is the residual bacteria cell numbers (biomass) after the simulation, as our goal is to design surfaces that optimally remove the biofilm under fluid flow and hence reduce bacteria cell numbers. The design variables \textbf{x} are subjected to a range of lower and upper bounds given in Section \ref{simulation}. The design variables $\mathbf{x} = \left[R_x, R_y, h, n, (M, T)\right]$ are the lower and upper radius of the cones, the height of the cones, and the number of cones on each side of the simulation box; and $(M, T)$ are the magnitude and time per vibration cycle, respectively. The magnitude and time are demarcated with brackets as they are not design variables in the growth and shear optimizations. The lower radius of each cone is larger than the upper radius, $R_x \geq R_y$, but both are non-zero and smaller than the maximum length per cone as a geometric constraint. The simulation parameters $\mathbf{p} = \left[\alpha, \xi, \mathbb{L}, \mathcal{T}, \mathcal{B}\right]$ are the shear rate, growth rate, geometric parameters, simulation iterations, and bacteria related coefficients, respectively. $\mathbb{L} = [L_x, L_y, L_z, L_\mathcal{S}, L_\mathcal{B}, ...]$ is the set of all parameters needed to set up the geometry of the active surfaces, and $\mathcal{T}$ and $\mathcal{B}$ control bacterial growth and removal at specific simulation steps with user-specified biological coefficients. Further details are given in Section \ref{simulation}.

BO consists of surrogate models built with Gaussian process regression for evaluating the space using Bayes statistics and an acquisition function. The acquisition function is used to construct a utility function from the model posterior that enables the next point to be evaluated\cite{acquisition_func}. The two components are introduced and explained in Sections \ref{gpr} and \ref{acquisition}.

\subsubsection{Gaussian Process Regression\label{gpr}}

Gaussian process regression (GPR) is a Bayesian statistical approach to approximate and model function(s). Considering our optimization problem, if the function is denoted as $y = f(\mathbf{x, p})$, where $f$ is evaluated at a collection of different sets of points: $\mathbf{x}_1, \mathbf{x}_2, ..., \mathbf{x}_k \in \mathbb{R}^d$, we can obtain the vector $[f(\mathbf{x}_1), ..., f(\mathbf{x}_k)]$ to construct a surrogate model for the design parameters with the correlated objectives. The vector is randomly drawn from a prior probability distribution, where GPR takes this prior distribution to be a multivariate normal with a particular mean vector and covariance matrix. Here, the mean vector and covariance matrix are constructed by evaluating the mean function $\mu_0$ and the covariance function $\Sigma_0$ at each pair of points $x_i$, $x_j$. The resulting prior distribution on the vector $[f(x_1),..., f(x_k)]$ is represented in the form of a normal distribution to construct the surrogate model\cite{bayes}\begin{equation}
    f(\mathbf{x}_{1:k}) \sim \mathcal{N}\left(\mu_0 (\mathbf{x}_{1:k}), \Sigma_0 (\mathbf{x}_{1:k}, \mathbf{x}_{1:k}))\right)\label{surrogate}
\end{equation}
where $\mathcal{N}(\cdot)$ denotes the normal distribution. The collection of input points is represented in compact notation: $1:k$ represents the range of $1,2,..., k$.
The surrogate model $f(\mathbf{x})$ on $1:k$ is represented as a probability distribution given in Equation (\ref{surrogate}). To update the model with new observations, such as after inferring the value of $f(\mathbf{x})$ at a new point $\bf x$, we let $k = l+1$ and $\mathbf{x}_k = \mathbf{x}$. The conditional distribution of $f(\mathbf{x})$ given observations $\mathbf{x}_{1:l}$ using Bayes' rule is
\begin{equation}
    \begin{aligned}
   f(\mathbf{x})| f(\mathbf{x}_{1:l}) &\sim \mathcal{N}(\mu_l (\mathbf{x}), \sigma_l^2 (\mathbf{x}))\\
    \mu_l (\mathbf{x}) &= \Sigma_0 (\mathbf{x}, \mathbf{x}_{1:l}) \Sigma_0 (\mathbf{x}_{1:l},\mathbf{x}_{1:l})^{-1} \left(f(\mathbf{x}_{1:l}) - \mu_0 (\mathbf{x}_{1:l})+\mu_0(\mathbf{x}) \right)\\
    \sigma_l^2 &= \Sigma_0 (\mathbf{x}, \mathbf{x}) - \Sigma_0 (\mathbf{x}, \mathbf{x}_{1:l})\Sigma_0 (\mathbf{x}_{1:l}, \mathbf{x}_{1:l})^{-1} \Sigma_0 (\mathbf{x}_{1:l}, \mathbf{x})
    \end{aligned}
\end{equation}where the posterior mean $\mu_l(\mathbf{x})$ is a weighted average between the prior $\mu_0(\mathbf{x})$ and the estimation from $f(\mathbf{x}_{1:l})$, where the weight applied depends on the kernel used. 

Here, we use the Gaussian kernel, hence the prior covariance is\cite{jap}\begin{equation}
\begin{aligned}
    \Sigma_0 (\mathbf{x}_i, \mathbf{x}_j) &= \sigma^2 R(\mathbf{x}_i, \mathbf{x}_j),\\
    R(\mathbf{x}_i, \mathbf{x}_j) &= \mathtt{exp} \left( \frac{1}{2} \sum_{m=1}^d \frac{\left( \mathbf{x}_{i, m} - \mathbf{x}_{j, m}\right)^2}{\theta_m^2}\right)\\
    \theta_m &= (\theta_1, \theta_2, ..., \theta_d)
\end{aligned}
\end{equation}where $\sigma^2$ is the overall variance parameter and $\theta_m$ is the correlation length scale parameter in dimension $m$ of the $d^{th}$ dimension of $\mathbf{x}$, which are all hyperparameters of GPR. $R(\mathbf{x}_i, \mathbf{x}_j)$ is the spatial correlation function. Our goal is to estimate the parameters $\sigma$ and $\theta_m$ that create the surrogate model given the training data $[y_k = \mathcal{N}_{BC(k)},\ \mathbf{x}_k]$ at iteration $k$.

\subsubsection{Acquisition Function\label{acquisition}}

Given the training data $[y_k,\ \mathbf{x}_k]$, Equation (\ref{surrogate}) gives us the prior distribution $y_l \sim \mathcal{N}(\mu_0, \Sigma_0)$ as the surrogate. This prior and the given dataset induce a posterior: the acquisition function, denoted as $\mathcal{A}: \mathcal{X}\xrightarrow{} \mathbb{R}^+$, determines the point in $\mathcal{X}$ to be evaluated through the proxy optimization $\mathbf{x}_{\sf best} = \argmax_{\sf x}\mathcal{A}(\mathbf{x})$. The acquisition function depends on the previous observations, which can be represented as $\mathcal{A} = \mathcal{A}(\mathbf{x}; (\mathbf{x}_l, y_l), \theta)$. Taking our previous notation, the new observation is probed through the acquisition\cite{msde_nanoporous_material} \begin{equation}
    \mathbf{x}_{k}=\mathbf{x}_{l+1}= \argmax_{x\in \frac{\mathcal{X}}{\mathcal{X}_l}} \mathcal{A}\left( \mathbf{x};(\mathbf{x}_l, y_l), \theta_m\right)
\end{equation}where the input space contains the evaluation of design variables at $n$ points: $\mathcal{X}_l : = (\mathbf{x}_1, \mathbf{x}_2, ..., \mathbf{x}_l)$. In our case, $\mathcal{X}$ is acquired through running $n$ numbers of NUFEB simulations. We pick the GP Upper Confidence Bound (GP-UCB)\cite{why_ucb} as the acquisition function, exploiting the lower confidence bounds (in the case of minimizing the objective function) to construct the acquisition and minimize the regret. GP-UCB takes the form\cite{acquisition_func}\begin{equation}
    \mathcal{A}\left({\bf x}; (\mathbf{x}_l, y_l), \theta_m\right):= \mu_l \left(\mathbf{x}; (\mathbf{x}_l, y_l), \theta_m\right) + \kappa \sigma\left(\mathbf{x}; (\mathbf{x}_l, y_l), \theta_m\right)\label{acquisition_function_equation}
\end{equation} where $\kappa$ is a tunable parameter balancing exploitation and exploration when constructing the surrogate model. We take $\kappa=2.5$ as a default value in the model. Combining GPR and the acquisition function, the surrogate model can be constructed to approximate the minimum value in the design space. In our case, such BO methods are applied to obtain active surface typologies with minimal residual bacterial cells. The design space is a 4-dimensional space for topology optimization and a 6-dimensional space for combined vibration optimization. We randomly explore the design space for 10 steps for the initial surrogate modeling and then iterate for 90 steps based on Bayesian statistics to construct the full surrogate with 100 data points.


\subsection{Simulation Setup\label{simulation}}

\begin{figure}[htpb]
    \centering
    \includegraphics[scale=0.3]{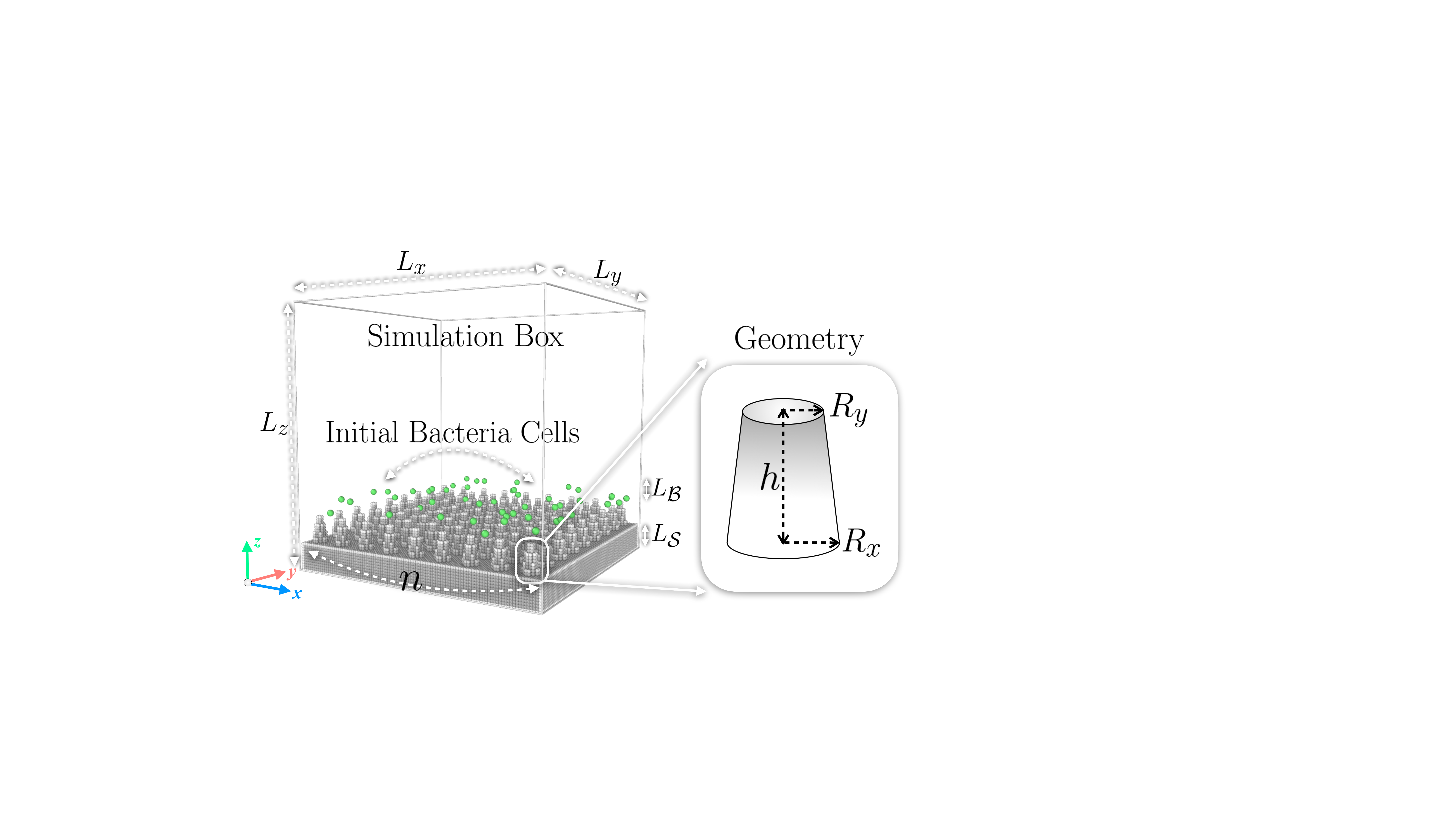}
    \caption{The numerical setup of the simulation box, implemented in {\tt LAMMPS}. A cubic simulation box with sizes of $L_x = L_y = L_z = 4\times 10^{-5}$ m. The geometry of the surface is controlled by four design parameters, $R_x$, $R_y$, $h$, and $n$, denoting the bottom and upper radii, the height of the cones, and the number of cones per box length, respectively. }
    \label{simulationbox}
\end{figure}

Our model (Figure \ref{simulationbox}) uses a cubic simulation box of lengths $4\times10^{-5}$ m with a substrate of height $L_\mathcal{S} = 4 \times 10^{-6}$ m. Above the substrate, we place cones with height $h$ as a design variable, which ranges between $[2 \times 10^{-6}\ {\rm m}, 4 \times 10^{-6}\ {\rm m}]$. The initial bacteria cells are placed above these cones for simulating biofilm growth, with a height of $L_\mathcal{B} = 2 \times 10^{-6}$ m. $n$ is the total number of cones per cubic side. Since the simulation box is cubic, the total number of cones should be $n \times n$. $n$ is an integer constrained in the range $[5, 10]$. For each $n$, the maximum value of the cone radii is $L_x/n$. The tunable range of the two radii are set as $[0.1,\ 0.9]\times \frac{L_x}{2n}$, corresponding to the  ``{\em geometric constraint}'' mentioned in Section \ref{workflow} herein. Since the geometric constraint, $R_x \geq R_y$ is assumed, the radii will be swapped if a larger $R_y$ is proposed by the optimization algorithm. We also apply vibrations to the substrate with magnitudes in the range $[4\times 10^{-7}\ {\rm m}, 2\times 10^{-6}\ {\rm m}]$ and time periods in the range $[ 10^{-5}\ {\rm s}, \ 1\ {\rm s}]$. 

Three different types of particles are involved in the simulation (Figure \ref{simulationbox}): the heterotrophs (HET), which can be interpreted as the bacteria cells; the extracellular polymeric substances (EPS); and the substrate, modeled as rigid particles. Note that for different loading, the simulation setup is mildly varied to tune the simulation. For instance, there are no vibration magnitude and time involved for pure growth and shear flow removal. Further details are to be explained in the following points.

To find the ideal surface topologies for effective biofilm removal, we optimize (1) the geometry of the nanosurface subjected to shear flow, or (2) both the geometric and vibration parameters. These two scenarios are modelled as follows:\begin{itemize}
    \item {\bf Biofilm growth and applied shear flow}. The substrate is rigidly fixed and bacteria cells are introduced right above the cones. Each side of the simulation box has fixed boundary conditions (FBC) such that the bacteria cells are removed if they exit the simulation box. To apply shear flow, from the equation of velocity of applied forces\cite{Li2019} we can deduce the equation controlling the shear rate as\begin{equation}
        \mathbf{F}_{f,i} = 6 \pi \mu r_i \mathbf{v}_r \xrightarrow{\alpha = \frac{d \mathbf{v}_r}{dt}} \alpha = \frac{dt}{6\pi\mu dr_i}\frac{d \mathbf{F}_{f,i}}{dt}\label{shear_equa}
    \end{equation}where $\mu$ is the dynamic viscosity, taken as $\mu = 0.001 \rm \ kg \cdot m^{-1}\cdot s^{-1}$ in our approach to model the flow of water\cite{water_density}, $\mathbf{v}_r$ is the local velocity of the particle, $\mathbf{F}_{f,i}$ is the shear force, and $\alpha$ is the applied shear rate, which is the time derivative of velocity. We assign $\xi_{HET} = 0.00028 \rm\ s^{-1}$ as the growth rate for heterotrophs\cite{Li2019} and $\alpha = 0.3\rm\ m\cdot s^{-2}$ as the shear rate. From Equation (\ref{equation_growth}) one can interpret $\xi$ as the change of bacterial mass, taking the form $m_i = \rho_i \frac{4}{3}\pi r_i^3$ for the $i^{th}$ particle. The density of EPS is $\rho_{EPS} = 30 \rm \ kg \cdot m^{-3}$\cite{Li2019, nufeb_frontier}. The density for the substrate is $\rho_{substrate} = 4410 \rm \ kg \cdot m^{-3}$, using Ti-6Al-4V\cite{ti_al_density} as our reference material considering the potential applications in additive manufacturing. The density for HET is $\rho_{HET} = 150 \rm \ kg \cdot m^{-3}$ based on previous experiments\cite{het_density}. 50 bacteria cells are randomly distributed above the cones initially and their growth is simulated for 200,000 s in real time, governed by Equations \ref{equation_growth}-\ref{relax_equilibrium_mechanical}. After the growth, shear flow is applied in the box for another 200,000 s as governed by Equations \ref{cfd1}, \ref{cfd2}, \& \ref{shear_equa}.
    
    \item {\bf Vibration induced biofilm detachment}. To allow for vertical vibrations of the substrate, we locate the substrate at an initial height $L_\mathcal{I} = M$ while maintaining FBC. For lateral vibrations, periodic boundary conditions (PBC) are set in the Y-direction. According to the implementation in \texttt{LAMMPS}\cite{lammps}, the displacement of any bacteria cells $\bf X$ takes the form\begin{equation}
        \mathbf{X}(t) = \mathbf{X}_0 + M \sin \left( \frac{2\pi}{T} \delta\right)
    \end{equation}where $\mathbf{X} = [X, Y, Z]$ is the position vectors of each particle, $\mathbf{X}_0$ is the initial position vectors, with vibration magnitude $M$ and time $T$ following Equation (\ref{opt_problem_equation}). $\delta$ is the elapsed time. For both vertical and lateral vibrations, 500  bacteria cells are randomly distributed above the cones and allowed to grow for 20,000 s in real-time. Vibrations are then applied for another 10,000 s in real-time. 
\end{itemize}

\subsection{Automated Optimization Workflow\label{workflow}}

\begin{figure}[htpb]
    \centering
    \includegraphics[scale=0.3]{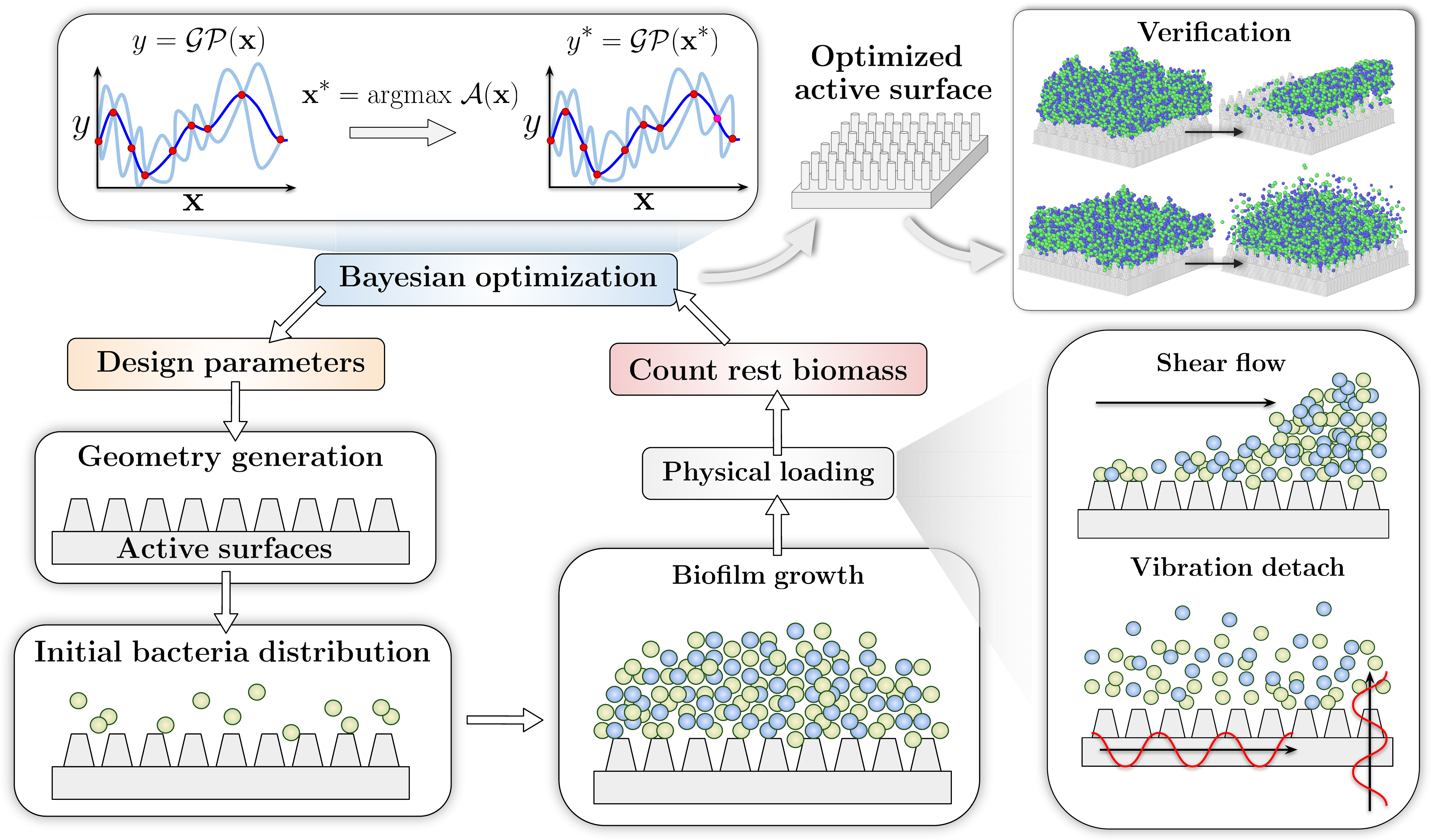}
    \caption{The schematic for the BO workflow for designing antimicrobial surfaces based on \texttt{LAMMPS} and \texttt{Python}. The optimization begins with randomly initiated geometries represented via the design parameters. The bacteria cells are initiated on top of the nanosurface and grown and removed via different physical loading. The remaining bacteria cells are the objective for the optimization. The optimized geometries are then verified through numerical simulations.}
    \label{fig2}
\end{figure}

Coupling the optimization process given in Equations \ref{surrogate}-\ref{acquisition_function_equation}, and the simulation processes given in Equations \ref{equation_growth}-\ref{cfd2} via Equation (\ref{opt_problem_equation}), we develop a general automated BO workflow enabled by \texttt{LAMMPS}-\texttt{Python} interface\cite{lammps} (Figure \ref{fig2}). The full optimization begins with generated design parameters implemented in \texttt{Python}. The design parameters are then translated into particle-represented geometries through NUFEB\cite{Li2019} implemented in \texttt{LAMMPS}. NUFEB simulations are then performed with the initial bacteria distribution, growth, and physical removal methods as described in Section \ref{simulation}. After performing the simulations, the residual bacteria cells for both HET and EPS are counted and passed to the optimization algorithm. 

Initially, 10 sets of data representing the geometry are randomly generated for building the raw surrogate model. The $11^{th}$ to $100^{th}$ geometries are then probed through the lower bound acquisition function using the same workflow for building the eventual surrogate, which we term as the metamodel. With this final metamodel, we can extract the optimized geometry for further simulations to verify and biomechanically rationalize why such surfaces are optimal for antimicrobial materials design.

\section{Results and Discussion\label{result}}

\subsection{Metamodels for Optimization\label{metamodel}}

\begin{figure}[htpb]
    \centering
    \includegraphics[scale=0.32]{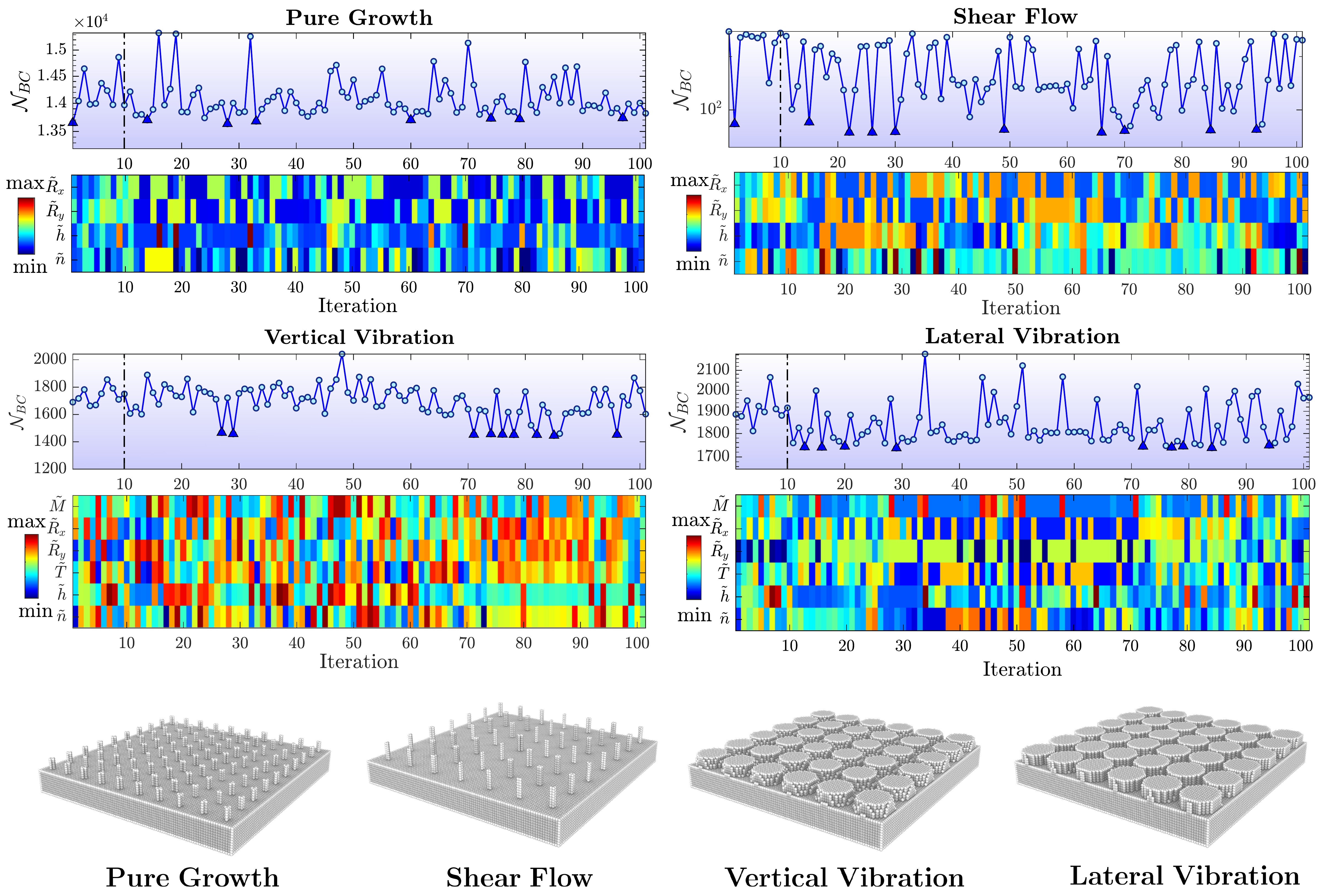}
    \caption{The design parametric matrices and the corresponding values of the objective function during the optimization process. Note that for better visualization the parametric matrices are normalized. Four subfigures indicate the optimization of the simulations based on four different physical methods to remove biofilm. The objective with lower biomass is marked in blue triangle dots for geometry extraction. The bottom figures stand for the optimized structures }
    \label{fig3}
\end{figure}

Figure \ref{fig3} shows metamodels generated from the optimization based on the four scenarios. From Figure \ref{fig3}, we observe that the lower objective values are distributed more uniformly throughout the iterations for the growth and shear flow cases, whereas for the vibration cases, the lower objectives seem to only exist under certain ``connected" steps where the corresponding design parameters exhibit similar values. 

\subsection{Geometry Representation\label{geometry}}


The lower value objectives marked in blue in Figure \ref{fig3} are extracted as these corresponding geometries seem to resist biofilm formation very well. 

Based on the mean values of the radii and heights corresponding to the most frequent cone numbers, the extracted optimized geometries for the four different scenarios are shown in subfigures \textbf{E1}, \textbf{E2}, \textbf{E3}, \textbf{E4}, respectively and the corresponding parameters are tabulated in Table \ref{optimized_parameters} in SI units. The four optimized geometries display exceedingly different characteristics: to purely minimize biofilm formation, the optimal geometry is $10\times10$ cones with relatively small radii and low height (subfigure \textbf{E1}). To efficiently remove biofilm under shear flow, the optimal geometry is taller cones with small radii and a larger distance between cones with cone numbers of $7\times7$ (subfigure \textbf{E2}). For both cases of applied vertical and lateral vibration (subfigure \textbf{E3} and \textbf{E4}), the optimized geometries have similar characteristics: total cones of $6\times6$ with short and thick cones. 

\begin{table}[htbp]
    \centering
    \begin{tabular}{c|cccc}
         & Pure Growth & Shear Flow & Vertical Vibration & Lateral Vibration \\\hline
        $R_x$ & $3.92\times10^{-7}$ [m] & $4.05\times10^{-7}$ [m] & $2.96\times10^{-6}$ [m] & $2.94\times10^{-6}$ [m] \\
        $R_y$ & $2.36\times10^{-7}$ [m] & $3.34\times10^{-7}$ [m] & $2.25\times10^{-6}$ [m] & $2.92\times10^{-6}$ [m]\\
        $h$ & $2.02\times10^{-6}$ [m] & $3.23\times10^{-6}$ [m] & $2.19\times10^{-6}$ [m] & $2.12\times10^{-6}$ [m]\\
        $n$ & 10 & 7 & 6 & 6\\
        $M$ & N/A & N/A & $4.7754\times 10^{-7}$ [m] & $4.9692\times 10^{-7}$ [m]\\
        $T$ & N/A & N/A & 0.1204 [s] & 0.1592 [s]\\\hline
    \end{tabular}
    \caption{The final optimized geometric and loading design parameters corresponding to Equation (\ref{opt_problem_equation}).}
    \label{optimized_parameters}
\end{table}

Observing both subfigures \textbf{E1} and \textbf{E2}, there are common characteristics of having small radii. As proposed by Hizal et al.\cite{acs_german_nanoengineer}, reduced adhesion is crucial for biofilm removal. Hence, we propose that geometric features such as thin ``pillar-like" shapes reduce the contact area between the biofilm and the substrate, as the reduced adhesion seems to both resist biofilm growth and promote shear flow removal. For both vibration scenarios, the resultant geometries are thick and short cylinders of large radii with fewer cones. A larger contact area is needed to transmit the vibration energy for biofilm removal. To elucidate these mechanisms in detail, we perform further IBM simulations and analyses of these optimal surface geometries. 

\subsection{Optimization Verification and the Biomechanics\label{biomechanics}}

\begin{figure}[htpb]
    \centering
    \includegraphics[scale=0.25]{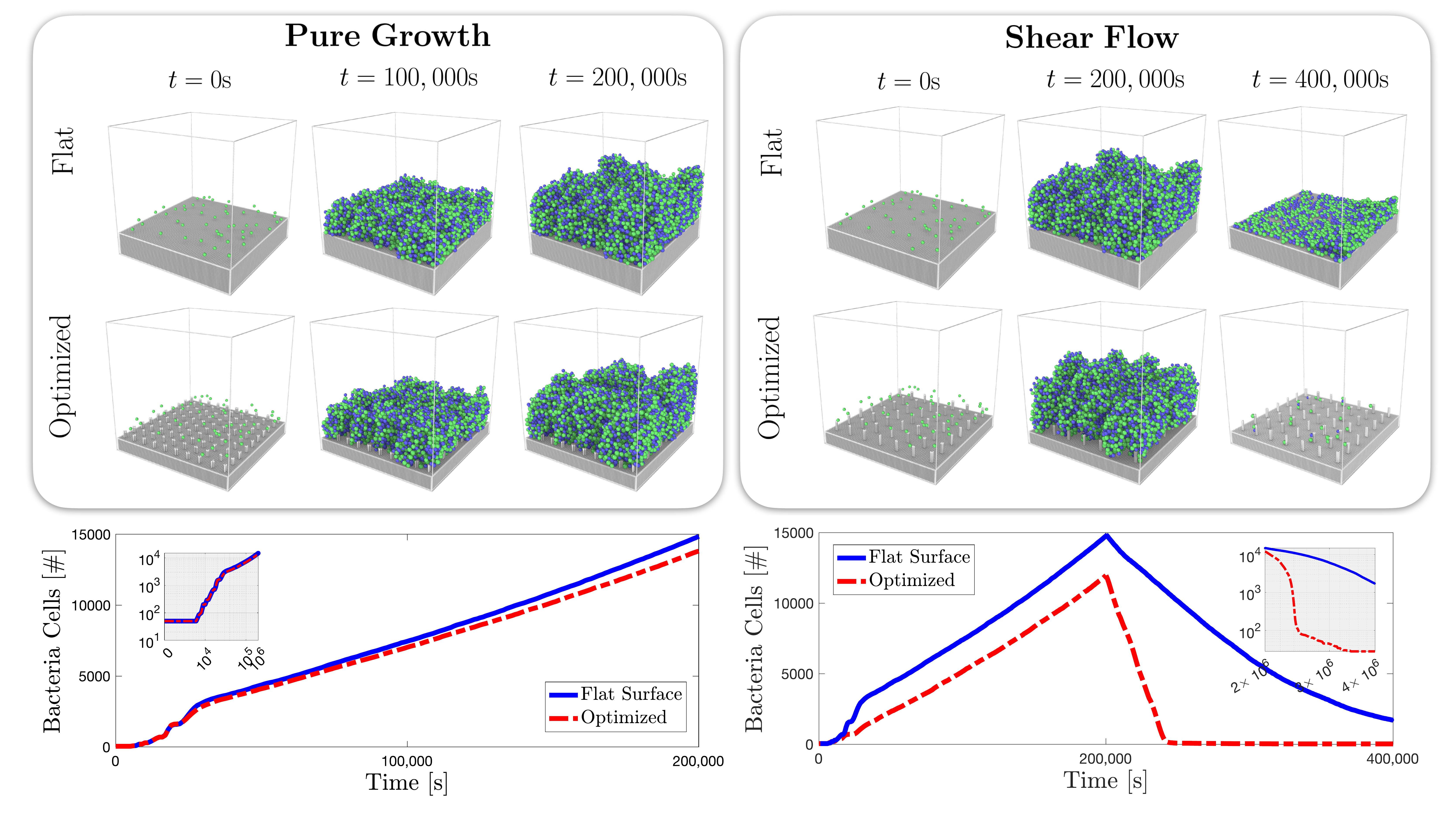}
    \caption{The performance of the optimized geometries is compared with a perfectly flat surface for resisting biofilm growth and shear flow removal. The top row sub-figures are simulation snapshots of the biofilm growth and removal corresponding to real-time. The bottom row is the plot of the changes in the bacteria numbers with respect to time during the simulation, thus highlighting the difference in biofilm removal efficiencies between the flat and optimized surfaces. The inset figure in the left-bottom sub-figure is the double logarithmic plot during the growing process, showing that the optimized active surface does not strongly alter the biofilm growing process in the simulations. The inset figure in the right-bottom sub-figure is the double logarithmic plot during the ``shear-off" process, showing that the optimized surface evidently improves the biofilm shearing removal. }
    \label{fig5}
\end{figure}

To analyze the mechanisms of action for the optimized geometries, numerical simulations are compared for the four scenarios. The optimized active surfaces are compared with flat surfaces for just the biofilm growth and the shear flow removal shown in Figure \ref{fig5}. 

Figure \ref{fig5} shows that the optimized active surface for purely resisting biofilm growth does not exhibit evident improvement compared with a perfectly flat surface, as from both the plot and visualization the optimized one does not greatly reduce the total bacteria cells. We can conclude that altering the surface topologies alone is insufficient for reducing the growth and formation of biofilm, especially since no other chemical effects are present, such as surface charges. However, when subjected to  shear flow, there is obvious biofilm reduction on the optimized surface compared with the flat one. We can hence contend that the optimization works well on surfaces designed for shear flow-induced biofilm removal as a secondary mode of action is needed to exploit the reduced adhesion of the biofilm.

The biofilm removal efficiency can be quantified by calculating the ratio of removed bacteria cells to the original bacteria cells before the biofilm growth:\begin{equation}
    \eta = \frac{ \mathcal{N}_{BC}(t_{G}) - \mathcal{N}_{BC}(t_{R})}{\mathcal{N}_{BC}(t_{G})}\label{biofilm_efficiency}
\end{equation}where $\mathcal{N}_{BC}(t_{R})$ is the bacteria cell number at the end of the simulation and $\mathcal{N}_{BC}(t_{G})$ is the bacteria cell number right after the end of the initial growth period of the biofilm. For the case of pure growth, we found that the optimized surface reduces bacteria by 6.82\% compared to the flat surface. Under shear flow, the biofilm removal efficiencies for the optimized and flat surfaces are 99.77\% and 88.5\%, respectively, suggesting an improvement of 11.27\% on biofilm removal efficiency. 

\begin{figure}[htpb]
    \centering
    \includegraphics[scale=0.23]{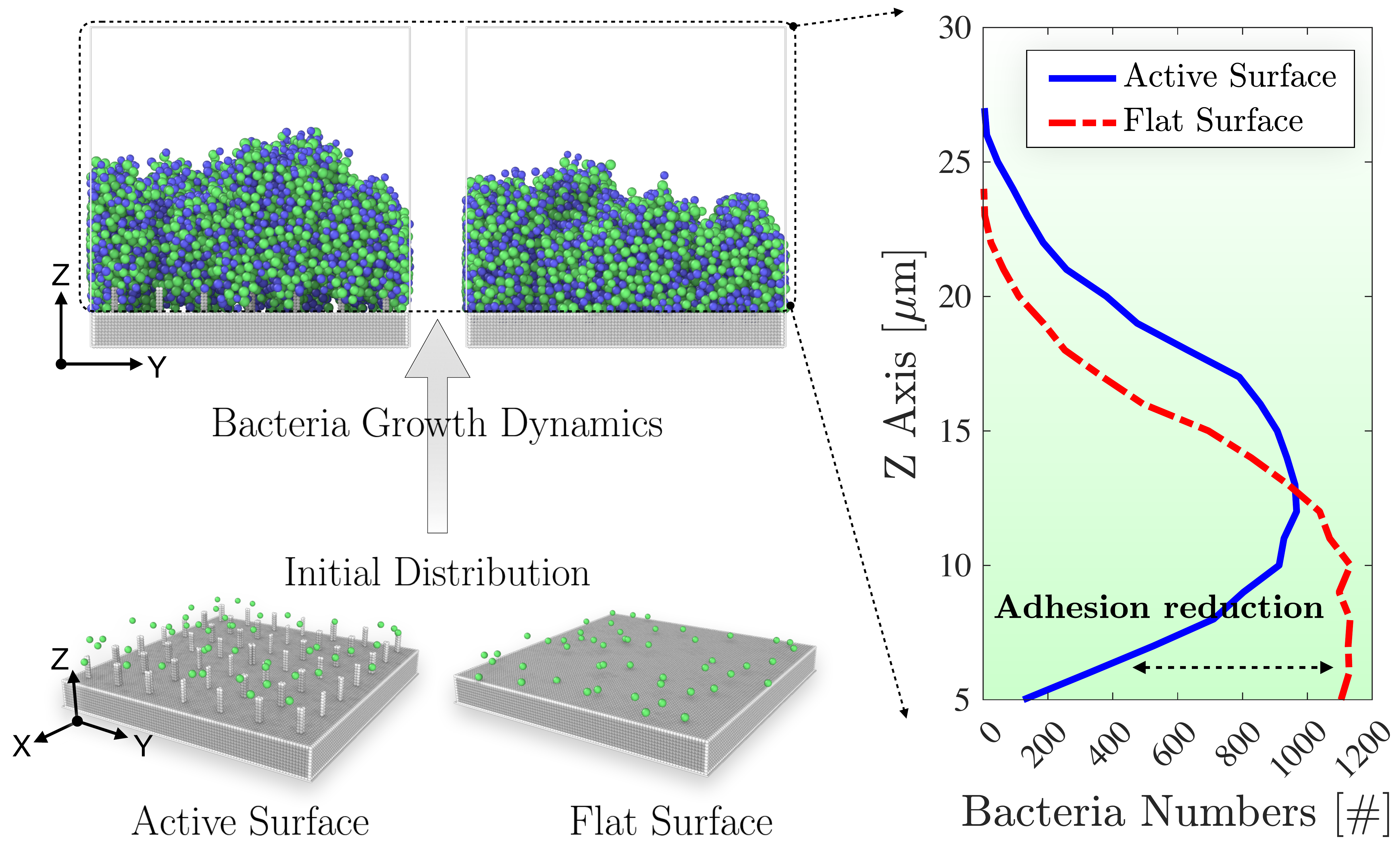}
    \caption{Schematic illustration for visualization of bacteria count with regards to height for visualizing the adhesion effects. The right diagram illustrates how the bacteria number distributes along the Z axis, where the blue and red lines stand for the flat and active surfaces, respectively. Note that the dashed red line denotes a ``cutoff" to indicate that the above area does not contain much bacteria, whereas the non-smooth bacteria number decrease is caused by the relatively low fidelity sampling.}
    \label{adhesion_explain_fig}
\end{figure}

Figure \ref{adhesion_explain_fig} shows how the bacteria number distribution along the Z axis to illustrate our proposed explanation of reduced adhesion of active surfaces leads to more efficient biofilm removals. In the right sub-figure, It can be observed that for the active surface, the biofilm clustered at a much higher location above $Z = 10\mu$m (blue) compared with the flat surface below $Z = 10\mu$m (red). Moving closer to the substrate surface ($Z=5\mu$m), the bacteria decreases drastically for the active surface yet compared to the flat surface. This phenomenon verifies and visualizes our proposition that the active surface topology can reduce the adhesion between the biofilms and their attaching surfaces.

\begin{figure}[htpb]
    \centering
    \includegraphics[scale=0.335]{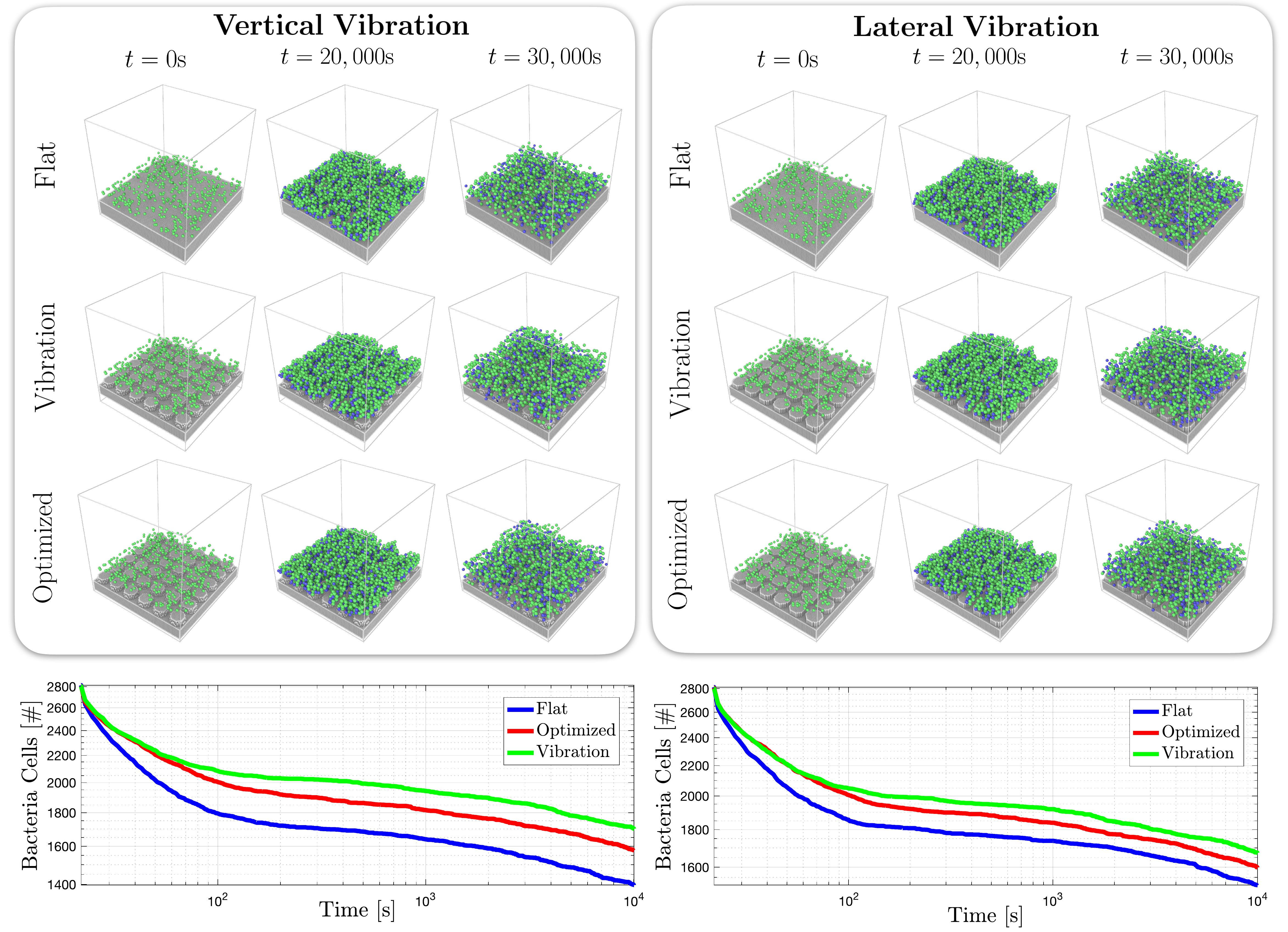}
    \caption{The numerical verification compares the optimized geometry with the optimized frequency properties and the benchmark coefficients, with the flat surface for vibrational biofilm removal. The upper figures show the snapshots of the simulation for removing biofilm using different vibration loadings, comparing the optimized geometry and loadings with alternating the geometry (flat surface) and vibration properties.}
    \label{fig6}
\end{figure}

Since both the geometries and vibration properties are optimized under different vibration-induced biofilm removal, the optimized scenarios are compared with two benchmark numerical experiments: (1) fixing the optimized vibration loading and replacing the geometry with a flat surface; (2) fixing the optimized active surface and alter the vibration loading. The comparison numerical simulations are shown in Figure \ref{fig6}, where ``Flat" indicates we hold the vibration loading and alter the geometry to the flat surface and ``Vibration" indicate we hold the geometry yet alter the vibration loading.

When subjected to vibration, Figure \ref{fig6} verifies our hypothesis that a flat surface will exhibit better biofilm removal effects as a larger contact area increases the transmission of vibrational energy from the substrate to the biofilm. However, the optimized vibration loading in Table \ref{optimized_parameters} seems to be counter-intuitive: one may expect a robust vibration, i.e., larger magnitude and higher frequency, to be more efficient in removing biofilm. Yet, the optimal conditions call for a smaller magnitude with larger time periods than the lower bound of $10^{-5}$ seconds. We thence apply the smallest time period with the largest vibration magnitude for comparison experiments, indicated as ``Vibration" in Figure \ref{fig6}. The lower subfigure also suggests that this ``extreme" condition does not exhibit a better biofilm removal effect, comparing the green and red lines. Applying Equation (\ref{biofilm_efficiency}) we obtain the biofilm removal efficiency in the vertical vibration case in 10,000 seconds for flat surface, altered vibration loading, and optimized scenario are 50.14\%, 39.26\%, and 43.65\%, respectively. The biofilm removal efficiency in the lateral vibration case in 10,000 seconds for flat surface, altered vibration loading, and optimized scenario are 45.96\%, 40.21\%, and 42.89\%, respectively. We can further contend that for vertical vibration case, the optimized scenario reduces 6.49\% compared with flat surface yet increase 4.39\% compared with altering the vibration loading, on biofilm removal efficiency; for lateral vibration case, the optimized scenario reduces 3.07\% compared with flat surface yet increase 2.68\% compared with altering the vibration loading, on biofilm removal efficiency. In fine, one may conclude that in the optimization for the vibration cases the algorithm proposes geometries that are similar or mimic flat surfaces, under the specific time period and magnitude loading. Interestingly, the counter-intuitive results of the optimized vibration loading may inspire future biofilm control strategies.

\section{Conclusion and Outlook\label{conclusion}}

In this study, we couple BO with individual-based models and simulations to propose an  automated machine-learned topological design workflow for designing antimicrobial active surfaces from sparse data points. The metamodels are constructed by collecting data from 100 simulations. The optimized workflow is applied to multiple case studies of purely resisting biofilm formation, removal of biofilm by applying shear flow, and detaching biofilm using vertical and lateral vibrations. We optimized the corresponding active surfaces under these different physical environments. The algorithms proposed four different geometries with corresponding vibration loading parameters. For purely resisting biofilm growth, the optimized active surface reduces biofilm formation by 6.82\%. Under shear flow, 88.50\% of the biofilm is removed from a perfectly flat surface,  compared to the 99.77\% removal rate from the optimized active surface, thus signifying improved efficiency of 11.27\%. When subjected to vertical vibration, the optimized scenario reduces 6.49\% compared with a flat surface yet increases 4.39\% compared with altering the vibration loading, on biofilm removal efficiency. For the lateral vibration case, the optimized scenario reduces 3.07\% compared with a flat surface yet increases 2.68\% compared with altering the vibration loading, on biofilm removal efficiency. 

We further found that under pure growth or applied shear flow, the optimal designs with lower objective values are more uniformly distributed during the iterative process. However, the optimal designs for both cases of vertical and lateral vibrations are more densely clustered in certain iterations. The optimized geometries are extracted from all the selected optimal design parameters by first selecting the target cone numbers and averaging the radii and heights. For purely resisting biofilm growth, the optimal geometry consisted of large numbers of thin and short cones. Under applied shear flow, the optimized geometry exhibited sparse cones with thin and tall pillar-like cones. For both vibration cases, the optimized geometries all display short and thick cylinder-shaped cones with fewer cones in the simulation cell, which can be interpreted as approximating a flat surface. Interestingly, the optimized vibration loading shows low vibration magnitudes with vibration time periods on the order of 0.15s, which is counter-intuitive. 

In brief, our study proposes methods to rapidly design antimicrobial topographies based on physical environments using simulations and optimization algorithms, enabling the machine-learned design of engineered antifouling surfaces. Our study is intended to inspire further investigations on (1) biofilm control strategies, both experimental and numerical, considering shear flow, vibration, and other possible methods; and (2) simulation-enabled machine-learned biomaterials design.

\section*{Data Availability}

All the data and code used in this paper will be made publicly available at \url{https://github.com/hanfengzhai/PyLAMDO}.

\section*{Conflicts of Interests}

There are no conflicts to declare.

\section*{Acknowledgement}

J.Y. acknowledges support from the US National
Science Foundation (Grant No. 2038057) and the Cornell University faculty startup grant. The authors also acknowledge the computational resources provided by the XSEDE
program under Grant TG-MAT200004 and TG-BIO210063, and
the computational resources provided by the G2 cluster from
Cornell University. H.Z. gratefully acknowledges the fruitful discussions with Profs. Derek Warner and Sadaf Sobhani at Cornell University.


\bibliography{new}

\clearpage

\end{document}